\documentclass[5p]{elsarticle}
\journal{Physics Letters B}
\usepackage{graphicx}
\usepackage{amsfonts}
\usepackage{amsmath,amssymb,tensor}
\usepackage[usenames]{color}
\usepackage{upgreek}
\usepackage{bm}
\usepackage{dcolumn}
\usepackage{epsfig}
\usepackage{graphics}
\usepackage{subfigure}
\usepackage{wrapfig}
\usepackage[latin1]{inputenc}
\usepackage{latexsym}
\usepackage{rotating}
\usepackage{float}
\usepackage{hyperref}
\usepackage{xspace} % Sensible space treatment at en

\newcommand{\dd}[1]{\mathrm{d}#1\,}

\newcommand{\Int}{{\mbox{\tiny int}}}
\newcommand{\EH}{{\mbox{\tiny EH}}}
\newcommand{\GR}{{\mbox{\tiny GR}}}

\begin{document}
\begin{frontmatter}
\title{Gravitational-Wave Mediated Preheating}

\author[dartmouth]{Stephon Alexander}
\author[dartmouth]{Sam Cormack\corref{cor1}}
\ead{samuel.c.cormack.gr@dartmouth.edu}
\author[fudan]{Antonino Marcian\`o}
\author[montana,kavli]{Nicol\'as Yunes}

\cortext[cor1]{Corresponding author}

\address[dartmouth]{Center for Cosmic Origins and Department of Physics and Astronomy, Dartmouth College, Hanover, New Hampshire 03755, USA}
\address[fudan]{Center for Field Theory and Particle Physics \& Department of Physics, Fudan University, 200433 Shanghai, China}
\address[montana]{Department of Physics, Montana State University, Bozeman, MT 59717, USA}
\address[kavli]{Kavli Institute for Theoretical Physics, University of California, Santa Barbara, CA 93106, USA}

%%%%%%%%%%%%%%%%%%%%%%%%%%%%%%%%%%%
\begin{abstract}
\noindent 
We propose a new preheating mechanism through the coupling of the gravitational field to both the inflaton and matter fields, without direct inflaton-matter couplings. The inflaton transfers power to the matter fields through interactions with gravitational waves, which are exponentially enhanced due to an inflation-graviton coupling. One such coupling is the product of the inflaton to the Pontryagin density, as in dynamical Chern-Simons gravity. The energy scales involved are constrained by requiring that preheating happens fast during matter domination. 

\end{abstract}

\begin{keyword}
arXiv: 1405.4288

preheating \sep gravitational waves \sep Chern-Simons

\PACS 98.80.-k \sep 98.80.Qc \sep 98.80.Cq
%Cosmology, 98.80.-k
%Quantum cosmology, 98.80.Qc
%Inflationary universe, 98.80.Cq
\end{keyword}
\end{frontmatter}

%%%%%%%%%%%%%%%%%%%%%%%%%%%%%%%%%%%

%--------------------------------------------------
\section{Introduction}
Inflation is the paradigm wherein the universe undergoes exponential expansion, resolving the horizon, entropy and structure formation problems that plague the standard big bang scenario. It is usually believed that inflation ends once the inflaton reaches the bottom of its potential, at which point a new mechanism must act to transfer the inflaton's kinetic energy into a process that leads to particle creation. One such mechanism is \emph{preheating}~\cite{Traschen:1990,Shtanov:1995,Kofman:1994,Kofman:1997}: the inflaton enters a phase of parametric resonance, as it oscillates around the minimum of its potential, and through a direct matter-inflaton coupling, it leads to particle creation.   There is a large number of possible direct couplings between the inflaton and the standard model, and one must usually pick one somewhat arbitrarily.

But what if the coupling between the inflaton and the matter fields were \emph{indirect}? Because of the equivalence principle, the graviton will interact with all matter fields and its coupling will be non-arbitrary. Let us then consider the inflaton coupling to matter fields through a graviton intermediary. That is, consider the inflaton at the end of inflation depositing its kinetic energy in the graviton, which due to a direct graviton-inflaton coupling becomes parametrically excited, and then deposits its energy in the matter fields. This can happen if there are new couplings between the inflation field and the graviton so that when the inflaton oscillates it causes resonances in the graviton's modulation. We will show that such couplings are possible via Chern-Simons modified general relativity.  As we shall see, no direct inflaton-matter coupling will be necessary to obtain preheating in this gravitational-wave mediated scenario. 

\section{Action and Evolution Equations}
Many inflationary paradigms exist, but for concreteness consider the following Chern-Simon extension to Chaotic Inflation \cite{Linde:1983gd,Kaloper:2008fb}
\allowdisplaybreaks[4]
\begin{align}
\label{action}
S &= \int\dd{^4}x\sqrt{-g}\left[\mathcal{L}_{\EH} + \mathcal{L}_{\Int}  + \mathcal{L}_\phi + \mathcal{L}_\chi\right]\,,\\
\mathcal{L}_{\EH} &= \frac{R}{16\pi G}\,, \qquad
\mathcal{L}_{\Int} =  \frac{\alpha}{4}\phi R\widetilde{R}\,, \\
\mathcal{L}_{\phi} &= -\frac{1}{2}\left(g\indices{^{\mu\nu}}\partial_\mu \phi\partial_\nu \phi + m_\phi^2 \phi^2 \right)\,,
\end{align}
where $\phi$ is the inflaton, $\chi$ is the matter field, $R$ is the Ricci scalar associated with the metric $g_{\mu \nu}$, $R \widetilde{R}$ is the Pontryagin density, ie.~the contraction of the Riemann tensor with its dual, and $\alpha$ is a coupling constant with dimensions of inverse mass (we work here in natural units $c = 1 = h$). Except for the interaction term ${\cal{L}}_{\Int}$, Eq.~\eqref{action} is just a simple model for inflation with a quadratic inflaton potential, arising from a Taylor expansion about its minimum. 

Many possible graviton-inflaton couplings could be considered, but the one presented above, $\mathcal{L}_{\Int}$, is well-motivated. Such a coupling arises naturally in a variety of frameworks: (i) in heterotic string theories upon 4-dimensional compactification and a low-energy expansion~\cite{Alexander:2006,Alexander:2004xd}; (ii) in loop quantum gravity when the Barbero Immirzi parameter is promoted to a field and coupled to fermions~\cite{Taveras:2008yf,Mercuri:2009zt}; (iii) in effective field theories of inflation~\cite{Weinberg:2008hq}; (iv) in dynamical Chern-Simons gravity~\cite{Alexander:2009}. Let us emphasize, however, that the gravitational wave-mediated preheating mechanism proposed here does not depend on this particular coupling. 

Regardless of the motivation, the theory described above should be considered \emph{effective}, a truncated low-energy expansion of a more fundamental theory that is thus valid only up to some energy cut-off $\Lambda$. The effective theory ceases to be a valid description when the interaction term $\mathcal{L}_{\Int}$ becomes comparable to the Einstein-Hilbert term $\mathcal{L}_{\EH}$. The former can be written as a total derivative if $\phi$ is constant. Therefore, to estimate its size we should first integrate by parts, moving a derivative from $R\widetilde{R}$ to $\phi$. The interaction term then becomes comparable to $\mathcal{L}_{\EH}$ when $\alpha \dot{\phi} \sim M_{p}^{2} (h_{0}f)^{-1}$, where $M_{p}=G^{-1/2}$ is the Planck mass, $f$ is the gravitational wave frequency and $h_{0}$ is the gravitational wave amplitude. Saturating $\alpha$ at $\Lambda^{-1}$, $\dot{\phi}$ at $HM_{p}$, $h_{0}$ at unity and $f$ at $H$, where $H$ is the Hubble parameter, one finds  $\Lambda = (H/M_{p})^{2} M_{p}$, which of course satisfies $\Lambda \ll M_{p}$. Another consequence of the truncation of the effective theory at this order is that the terms neglected in the expansion, such as $(\partial \phi)^{4}/\Lambda^{4}$, are indeed small and ignorable. A consequence of all of this is that the interaction term $\mathcal{L}_{\Int} $ acts as a small perturbation to whichever inflationary mechanism one wishes to consider, and thus, it does not spoil (or really affect) inflation, until inflation ends and the inflaton reaches the bottom of its potential.

Variation of the action in Eq.~\eqref{action} with respect to all degrees of freedom leads to the field equations~\cite{Alexander:2009}
\begin{align}
\label{modified-EEs}
G\indices{_{\mu\nu}} +16\pi G \; \alpha \; C\indices{_{\mu\nu}} &= 8\pi G \; T\indices{_{\mu\nu}}\,,
\\
\label{modified-field}
\square \phi - m_\phi^2 \phi + \frac{\alpha}{4}R\widetilde{R} &= 0\,,
\\
\square \chi - m_\chi^2 \chi &= 0\,,
\end{align}
where $\square$ is the curved wave-operator, $G_{\mu \nu}$ is the Einstein tensor, $T_{\mu\nu}$ is the sum of the stress-energy tensors of the $\phi$ and $\chi$ fields, and
\begin{equation}
C\indices{^{\mu\nu}} \equiv \nabla\indices{_\alpha}\phi\, \epsilon\indices{^{\alpha\beta\gamma(\mu}}\nabla\indices{_\gamma}R\indices{^{\nu)}_\beta} + \nabla\indices{_{(\alpha}}\nabla\indices{_{\beta)}}\phi \widetilde{R}\indices{^{\beta(\mu\nu)\alpha}}.
\end{equation}
with $\widetilde{R}_{\beta (\mu \nu) \alpha}$ the dual Riemann tensor with indices symmetrized.

%--------------------------------------------------
\section{Order Reduction and Perturbation Theory}
Let us expand the equation for the metric tensor and the inflaton about a fixed background: $g\indices{_{\mu\nu}} = \overline{g}\indices{_{\mu\nu}} + \lambda h\indices{_{\mu\nu}}$ and $\phi(t,\mathbf{x}) = \overline{\phi}(t) + \lambda\delta\phi(t,\mathbf{x})$, where $\lambda$ is an order-counting parameter. The background $\overline{g}\indices{_{\mu\nu}}$ and $\overline{\phi}(t)$ will be taken to be the flat Friedmann-Lema\^itre-Robertson-Walker (FLRW) metric and a homogeneous and isotropic background field respectively, while $h\indices{_{\mu\nu}}$ and $\delta\phi(t,\mathbf{x})$ are first-order perturbations. 

The FLRW metric satisfies the background field equations exactly for any homogeneous and isotropic background inflaton field. The Hubble parameter is sourced by the energy density and pressure of this background field and the matter fields (for reasons that will become clear later, we do not decompose $\chi$). The background inflaton field satisfies the homogeneous and isotropic wave equation on an FLRW background with a mass potential. The Pontryagin density does not contribute to the background evolution of the inflaton, because this quantity vanishes exactly when evaluated for any spherically symmetric metric. 

To first-order in $\lambda$, the equations for the metric tensor perturbation become~\cite{Choi:2000}
\begin{align}
\label{eq:gwEOM}
\overline{\square} h'{}\indices{_{ij}} &=\frac{16\pi G}{a}\epsilon\indices{_{(i}^{lm}}[(\ddot{\overline{\phi}}-H\dot{\overline{\phi}})\dot{h'{}}\indices{_{j)l}}+\dot{\overline{\phi}}\,\overline{\square} h'{}\indices{_{j)l}}]\indices{_{,m}}
\nonumber \\ 
&+16\pi G a^2 p_\chi h'{}\indices{_{ij}}\,,
\end{align}
while, neglecting scalar metric perturbations (we are looking at modes shorter than the Hubble scale), the equation for the inflaton perturbation is 
\begin{align}
\label{eq:delphiEOM}
\delta\ddot{\phi} &+3H\delta\dot\phi - \frac{1}{a^2}\overline{\nabla}^2\delta\phi = -m_\phi^2\delta\phi\,,
\end{align}
where $\overline{\nabla}^2$ and $\overline{\square}$ are the Laplacian and wave operators in a homogeneous and isotropic FLRW background, $p_{\chi}$ is the pressure of the $\chi$ field and $h'{}\indices{_{ij}} \equiv a^{-2}h\indices{_{ij}}$. Notice again that the Pontryagin density does not enter the evolution equation of the inflaton perturbation, since it also vanishes identically to linear order in $\lambda$. 

We can simplify the evolution equation for the metric perturbation through order reduction. As discussed in~\cite{Yunes:2009hc,Garfinkle:2010zx,Ayzenberg:2013wua}, we decompose the metric perturbation into a general relativistic piece $h_{\mu \nu}^{\GR}$ and a deformation $\delta h_{\mu \nu}$, namely $h_{\mu \nu} = h_{\mu \nu}^{\GR} + \alpha^{2} \delta h_{\mu \nu}$. Note that the deformation is proportional to $\alpha^{2}$ because Eq.~\eqref{modified-EEs} is proportional to $\alpha C_{\mu \nu}$, $C_{\mu \nu}$ is proportional to $\phi$, and $\phi$ is proportional to $\alpha$ due to Eq.~\eqref{modified-field}. Using this decomposition, we can order reduce Eq.~\eqref{eq:gwEOM}: the left-hand side is proportional to $\overline{\square} \delta h_{\mu \nu}$, while the right-hand side is proportional to a differential operator acting on $h_{\mu \nu}^{\GR}$. This differential operator will contain one term of the form $\overline{\square}$, which automatically vanishes when acting on $h_{\mu \nu}^{\GR}$ because $R_{\mu \nu}[h_{\alpha \beta}^{\GR}] = 0$.  Using this and going to the transverse-traceless (TT) gauge~\cite{Alexander:2008} and in the left/right-circular polarization basis for the gravitational wave perturbation, Eq.~\eqref{eq:gwEOM} becomes
\begin{equation}
\label{eq:hREOM}
\overline{\Box} h_R = i \frac{16\pi G}{a}\alpha(\ddot{\overline{\phi}}-H\dot{\overline{\phi}})\frac{\partial}{\partial z}\dot{h}_R + 16\pi Ga^2p_\chi h_R\,.
\end{equation}
The equation for $h_L$ can be obtained by taking $i \to -i$ and $h_{L/R} \to h_{R/L}$. The amplitudes $h_L$ and $h_R$ are defined by
\begin{equation}
h'{}\indices{_{ij}} = \frac{1}{\sqrt{2}}\left(\begin{array}{ccc}
-(h_L+h_R) & i(h_L-h_R) & 0\\
i(h_L-h_R) & (h_L+h_R) & 0 \\
0 & 0 & 0 \\
\end{array}\right).
\end{equation}
Notice that Eqs.~\eqref{eq:gwEOM} and~ \eqref{eq:delphiEOM} are not coupled and can thus be solved independently, once the evolution of the background fields is obtained.

Let us now discuss the evolution of the matter fields. We anticipate that the matter occupation number will be generated through parametric resonance, so even a small perturbation of size ${\cal{O}}(|h_{\mu\nu}|)$, may have a large effect. We therefore treat the matter field, $\chi$, exactly, without a perturbative decomposition, while the gravitational field is treated to first order in its perturbations, $h_{\mu\nu}$. We obtain the equation to first order in $h_{\mu\nu}$,
\begin{equation}
\ddot{\chi}+3H\dot{\chi} -\frac{1}{a^2}\overline{\nabla}^2\chi + h\indices{^{ij}}\partial_i\partial_j \chi = -m_\chi^2 \chi\,.
\label{eq:chiEOM}
\end{equation}
In the TT gauge and in a circular GW polarization basis, this equation becomes
\begin{equation}
\ddot{\chi} +3H\dot{\chi} - \frac{1}{a^2}\overline{\nabla}^2\chi -\frac{1}{\sqrt{2}a^2}\mathrm{Re}[h_L\partial_L^2+h_R\partial_R^2]\chi = -m_\chi^2\chi\,,
\label{eq:chiEOMLR}
\end{equation}
where $\mathrm{Re}[x]$ is the real part of $x$ and we have defined $\partial_{L,R} \equiv ({\partial}/{\partial x} \mp i{\partial}/{\partial y})/{\sqrt{2}}$.

%--------------------------------------------------
\section{Behavior of Solutions}
Before attempting to solve the equations of motion, let us make some approximations. On the one hand, as the $\phi$ field oscillates, it will amplify and cause the gravitational waves to modulate, which in turn will drive the production of $\chi$ particles. The latter will therefore occur on the timescale $\tau_\phi = 1/m_\phi$. On the other hand, the expansion of space is governed by the Hubble parameter and the scale factor changes on the timescale $\tau_H = 1/H$. When $\tau_\phi \ll \tau_H$, or equivalently $H \ll m_\phi$, preheating occurs much faster than the expansion of space and one is justified in setting $a=1$ and $H=0$. Using that the Hubble parameter for a simple quadratic potential satisfies $M_{p}^{2} H^{2} = \frac{4\pi}{3} (m_\phi\, \phi)^{2}$, the requirement $H \!\ll \! m_\phi$ implies $\phi_0 \ll M_p$, where $\phi_0$ is the value of $\phi$ at the end of inflation. Even though it works well, this approximation can not be claimed to be excellent, as we know that $\phi_0<M_p$. Nevertheless, the approximation is correct, since the timescale involved in the growth of the particle number is smaller than the oscillation time. 

With this approximation, the equation of motion for the background inflaton field greatly simplifies to a that of a simple harmonic oscillator with frequency $m_{\phi}$, so 
\begin{equation}
\label{eq:phiSolFlat}
\overline{\phi} = \phi_0\sin(m_\phi t+\delta)\,,
\end{equation}
where $\phi_{0}$ and $\delta$ are constants of integration, ie.~the amplitude of the background inflaton and a phase shift. 

The background inflaton sources the evolution of the metric perturbation through Eq.~\eqref{eq:hREOM}. Using the approximation described above, and transforming to the Fourier domain, this equation becomes
\begin{align}
\label{eq:htildeEOM}
\ddot{\tilde{h}}_{R} &= \gamma \sin(m_\phi t +\delta)k\dot{\tilde{h}}_{R}
-(k^2+16\pi Gp_\chi)\tilde{h}_{R}\,,
\end{align}
where the overhead tilde stands for the Fourier transform and we have defined
\begin{equation}
\gamma \equiv 16\pi G\alpha m_\phi^2\phi_0 = 16\pi \frac{\phi_0 m_\phi^2}{M_* M_p^2}\,,
\end{equation}
which controls the strength of the oscillatory anti-damping term. The above equation of motion admits an exact solution through a linear combination of Heun functions times an oscillatory term, which we confirm by numerically solving Eq.~(\ref{eq:htildeEOM}).  We could obtain a similar expression for the left-polarized mode, but we do not need to. As found in~\cite{Alexander:2006,Alexander:2008,Yunes:2010yf}, Eq.~\eqref{eq:hREOM} leads to exponential amplification/damping of right-/left-handed gravitational waves during the inflationary epoch, so by the end of inflation, the left-handed gravitational waves are negligible and can be neglected during preheating. The left/right asymmetry is controlled by the sign of the coupling constant $\alpha$. Here we take $\alpha$ to be positive, which causes the left-handed gravitational waves to be damped.

With the solution to the background inflaton and the metric perturbation, we can now solve for the evolution of the matter fields. Working in the Heisenberg picture, we promote the matter field to a quantum operator $\hat{\chi}(\mathbf{x},t)$ and expand the latter in Fourier modes with raising and lowering operators, as done routinely in quantum field theory~\cite{Peskin:1995ev}. Using the flat-space approximation described above, the evolution of the Fourier mode functions obeys
\begin{multline} 
\ddot{\widetilde{\chi}} +(k^2+m_\chi^2)\widetilde{\chi}\\
 = - \frac{1}{\sqrt{4\pi}} \int\dd{k_z'}\left((k_x+i k_y)^2\tilde{h}_{R}(k_z-k_z')+\mathrm{h.c.}\right)\widetilde{\chi}' \,,
\label{eq:modeFunctionEOM}
\end{multline}
where $k$ is the magnitude of the $3$-momentum $k^{i}$-vector and $\mathrm{h.c.}$ stands for Hermitian conjugate. 

Let us contrast our matter production equation with the usual way particle creation occurs via the Mathieu equation.  In that case, the matter fields obey the following evolution equation
\begin{equation}
\ddot{\widetilde{\chi}} +(k^2+m_\chi^2)\widetilde{\chi} = -g^{2}\phi_{0}^{2} \sin^{2}(m_{\phi} t) \; \chi\,, \\ 
\end{equation}
where $g$ is the coupling constant between the $\chi$ and $\phi$ fields. Parametric resonance occurs because of the purely temporal oscillations of the inflaton.  In our case, however, the graviton couples through spatial gradients of the inflaton, and thus, parametric resonance occurs due to non-linear mode couplings and temporal oscillations in the gravitational wave's source term.
  
%--------------------------------------------------

\section{Parametric Resonance and Particle Creation}
The evolution equations for the gravitational wave and the mode functions, Eqs.~\eqref{eq:htildeEOM} and~\eqref{eq:modeFunctionEOM}, are those of a parametric oscillator, ie.~a harmonic oscillator with parameters, like the damping coefficient or the natural frequency, that oscillate in time at some other frequency. For example, the damping coefficient in Eq.~\eqref{eq:htildeEOM} is a function of time with frequency $m_{\phi}$, while the natural frequency in Eq.~\eqref{eq:modeFunctionEOM} is also an oscillatory function of time.

\begin{figure*}[thb]
\begin{center}
\includegraphics[width=\columnwidth,clip=true]{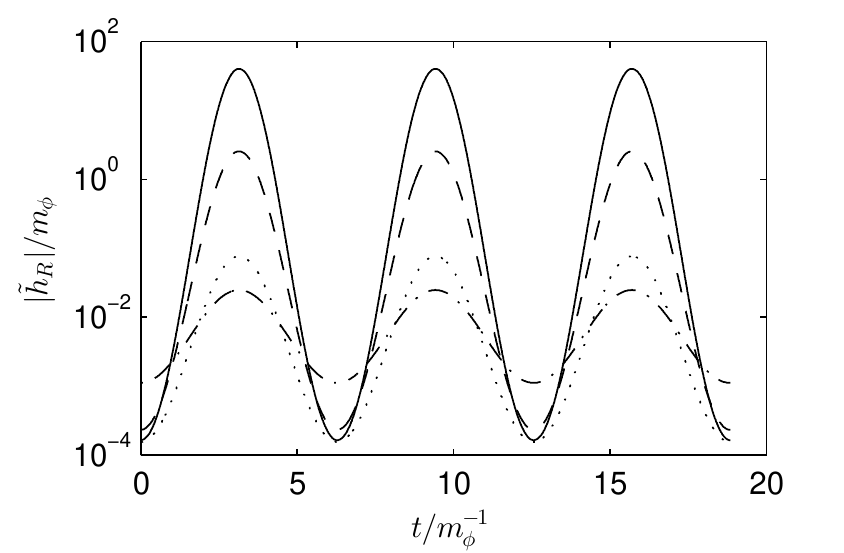}
\includegraphics[width=\columnwidth,clip=true]{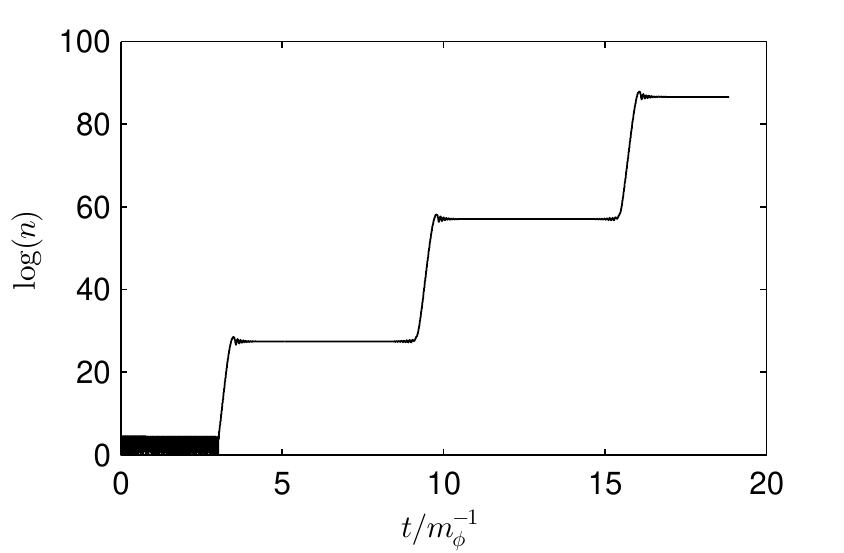}
\caption{\label{fig:hAmplitude} Left: Fourier transform of $h_{R}$ as a function of time in $m_{\phi}^{-1}$ units, for $k$ values of $50m_\phi$ (dashed dotted), $100m_\phi$ (dotted), $150m_\phi$ (dashed) and $200m_\phi$ (solid). We here choose $\gamma=0.062$, as this is the smallest value of $\gamma$ for which particle production occurs when $k=200m_\phi$ is the largest wavenumber used. We assume an initial scale-invariant power-spectrum for $\tilde{h}_{R}$. Right: Total particle number for $\chi$, calculated from the $\tilde{h}_{R}$ in the left-panel and summing over all wavenumbers. Most of the contribution to the sum comes from the larget wavenumbers, i.e.\ $\chi$ particles with large momenta are preferentially produced.  Calculations have been done for $q_z$ from $m_\phi$ to $100\, m_\phi$, $q_x=0$, $q_y=100m_\phi$.}
\end{center}
\end{figure*}

For the oscillatory damping to have an effect in the evolution of the metric perturbation, 
the quantity $\gamma k / m_\phi$ must be close to unity. The quantity $\gamma k$ is compared to $m_\phi$ since the damping must occur on timescales comparable to the oscillation of $\phi$. This implies that the energy scale $M_*=1/\alpha$ should satisfy
\begin{align}
\alpha\, \phi_0 &\sim \frac{M_p^2}{m_\phi k}.
\label{eq:conditionGamma}
\end{align}
We can compare this with the condition that $\mathcal{L}_\Int$ remains small compared to $\mathcal{L}_\EH$. By taking $\dot{\phi}\sim m_\phi \phi_0$ and identifying $f$ and $k$ we can write this condition as
\begin{equation}
\alpha \, \phi_0 \ll \frac{M_p^2}{m_\phi k h_0}
\label{eq:conditionCutoff}
\end{equation}
We can satisfy this equation if the gravitational wave amplitude satisfies $h_{0} \ll 1$, which is simply equivalent to the condition that $h_{\mu\nu}$ is a small metric perturbation.

We should also address the possible appearance of ghost instabilities in the theory. The analysis of ghost instabilities in Chern-Simons gravity has been performed by Dyda et al.\ \cite{Dyda:2012}. They find that at the linearized level, a ghost instability coming from the third derivative term in equation \eqref{eq:gwEOM} is only present if the momentum cutoff for the theory is greater than a mass scale they call the Chern-Simons mass, defined as $m_\mathrm{CS} = M_p^2/8\pi\alpha\dot{\phi}$. In our work, $\dot{\phi}\leq m_\phi \phi_0$, so, in terms of our parameter $\gamma$, we have $m_\mathrm{CS}\geq (2/\gamma)m_\phi$. 

In our numerical solution we use a cutoff of $\Lambda =200m_\phi$ while our Chern-Simons mass is $m_\mathrm{CS}\geq 32m_\phi$. If this situation persists to the present day then Dyda et al.\ show that too many photons will be produced by vacuum decay. However, the Chern-Simons mass  is not constant and, in our case, the situation $\Lambda>m_\mathrm{CS}$ occurs only during preheating. During slow roll inflation, the value of $\dot{\phi}$ will generically be smaller than during preheating, and the Chern-Simons mass will hence be greater. By the end of preheating, $\dot{\phi}\rightarrow 0$ so $m_\mathrm{CS}\rightarrow \infty$ and we return to the regime of standard general relativity. The ghost instability is therefore present only during the very brief period of preheating and possibly the very end of inflation. 

The presence of the ghost during preheating will mean that the vacuum is unstable to decay to light particles. Instability of the vacuum in the presence of a large classical field is not necessarily a fundamental problem. For instance, the Schwinger effect can be thought of as an instability of the vacuum in the presence of a large electric field. In our case the classical field is $\phi$ and the strength of the instability is controlled by $\dot{\phi}$. 

We have checked that vacuum decay will not be the dominant particle production effect in our scenario, and this provides an additional constraint on the energy scales of the model. The vacuum decay rate to photons is given by \cite{Dyda:2012}
\begin{equation}
\Gamma \sim \frac{m_\mathrm{CS}\Lambda^5}{M_p^2}.
\label{eq:vacuumDecay}
\end{equation}
The duration of preheating is of the order $10m_\phi^{-1}$, and let us assume that all of the produced photons have energy $\Lambda$. Then, using $\Lambda=200m_\phi$, our expression for $m_\mathrm{CS}$, and $\gamma\sim 10^{-1}$, we get an estimate for the energy density of photons from vacuum decay during preheating of
\begin{equation}
\rho_\mathrm{decay} \sim 10^{16}\frac{m_\phi^6}{M_p^2}.
\end{equation}
This should be much less than the energy initially stored in the inflaton field otherwise it will overwhelm any other particle production mechanism. The energy density in the inflaton field initially is $\rho_\phi \sim m_\phi^2 \phi_0^2$ and imposing $\rho_\mathrm{decay}<<\rho_\phi$ requires that
\begin{equation}
\frac{m_\phi^2}{M_p \phi_0} <<10^{-8}
\end{equation}
which does not contradict any of the other assumptions in our work. We will find that matter particles are produced exponentially via parametric resonance so that particle production from vacuum decay will be insignificant as long as the above condition holds.

Before presenting a solution to the evolution equations, let us define a good measure of particle production: the occupation number. Following~\cite{Kofman:1997}, this quantity in a given mode function is given by
\begin{equation}
n_{\vec{k}} = \frac{\omega_{\vec{k}}}{2}\left(\frac{|\dot{\widetilde{\chi}}|^2}{\omega_{\vec{k}}^2} + |\widetilde{\chi}|^2\right) - \frac{1}{2}\,,
\label{eq:particleNumber}
\end{equation}
where as usual $\omega_{\vec{k}}^2 = k^2 +m_\chi^2$.

We now have all the ingredients we need to explore particle production due to mediated parametric resonance. We solve Eqs.~\eqref{eq:htildeEOM} and \eqref{eq:modeFunctionEOM} numerically through the Dormand-Prince Runge-Kutta method \cite{Dormand:1980}. We ignore the back-reaction of $\chi$ on $\tilde{h}$, and thus first solve Eq.~\eqref{eq:htildeEOM} and then use this solution to solve Eq.~\eqref{eq:modeFunctionEOM}. The integral term in Eq.~\eqref{eq:modeFunctionEOM} is treated through the convolution theorem, using a fast Fourier transform. The numerical code is gridded in momentum from a minimum value of $m_\phi$ to $200m_\phi$. Any mode with momentum $k<m_\phi$ will not have time to oscillate much during the period of preheating, and thus can be ignored. The gravitational waves are assumed to start with a scale-invariant power-spectrum,
\begin{equation}
|\tilde{h}_R| = \frac{10\pi A_\mathrm{T}}{k^{3/2}}
\end{equation}
 but with large amplitude, $A_\mathrm{T}=1/20\pi$, since the right-handed waves are exponentially amplified during inflation. The parameter $A_\mathrm{T}$ is the standard gravitational wave spectrum amplitude. The $\chi$ field starts with Gaussian fluctuations with variance proportional to $1/k$.

Figure~\ref{fig:hAmplitude} shows our numerical results. The left panel shows the amplitude of the Fourier transform of $h_R$ as a function of time in units of $m_{\phi}^{-1}$ for different values of $k$. The right panel shows the corresponding growth in particle number, calculated by summing over wavenumber in Eq.~\eqref{eq:particleNumber}. Observe that there are large jumps in particle number when the amplitude of the gravitational waves becomes large. Energy must be conserved in the production of $\chi$ particles, and this energy must come from that contained in gravitational waves. The stored gravitational wave energy is proportional to the square of their amplitude, so the energy carried by them peaks just as the $\chi$ particles are being created. If we include the effect of $\chi$ back on $\widetilde{h}$, the peaks in amplitude will be much smaller. 

The number density, and hence the energy density, of $\chi$ particles increases by approximately ten orders of magnitude during each jump. We therefore conclude that this process is much more efficient than the vacuum decay mentioned earlier and that it will drain the energy from the inflaton field in only a few jumps. This supports our assumption that the expansion of the universe can be neglected during the preheating phase. We note that the amount of particle production from resonance increases with increasing momentum cutoff, similar to the vacuum decay rate (see equation \eqref{eq:vacuumDecay}). This suggests that the particle production is related to the ghost instability, though it is a nonlinear rather than perturbative effect.

%--------------------------------------------------
\section{Conclusion and Discussion} 
We have presented a gravitational wave-mediated preheating scenario where during its oscillatory phase, the inflaton deposits its energy to the graviton and the latter then excites matter production through parametric resonance. This mediated mechanism eliminates the need for  any {\it ad hoc} direct interaction between the inflation  and the matter fields, obviating issues of fine tuning the inflaton's interactions with those of the standard model.  In our case, the minimal coupling between the standard model and gravity is non-arbitrary.  We have provided a concrete realization of the mechanism in the context of Chaotic Inflation with a Chern-Simons coupling to the inflaton field.  These models are promising because they have already been implemented to generate lepton asymmetry during the inflationary epoch, through parity-violating gravitational waves~\cite{Alexander:2006}.  Furthermore, the notion that gravitational waves can generate preheating is generic so long as there are couplings of the inflaton to curvature invariants such as in Higgs-Inflation\cite{Bezrukov:2007ep}, which we leave for future investigations. We also plan to investigate the possibility of graviton loop induced corrections to the interactions in the theory which may enhance the preheating effect. 

%--------------------------------------------------
\section*{Acknowledgements}
We would like to thank Robert Brandenberger and Sean Carroll for useful discussions. NY~acknowledges support from NSF grant PHY-1114374 and the NSF CAREER Award PHY-1250636, as well as support provided by the National Aeronautics and Space Administration from grant NNX11AI49G, under sub-award 00001944. He also thanks KITP, and partial support from NSF grant PHY11-25915, for their hospitality during the completion of this work. SC thanks the Fulbright Foreign Student Program and Fulbright New Zealand for their support.

\section*{References}
\bibliographystyle{elsarticle-num}
\bibliography{paper_PLB_v2}
\end{document}